# FOCAL PROPERTIES OF PLANAR CURVILINEAR MIRRORS APPLIED TO HYDRODYNAMIC SOLITON ANALYSIS

**Germán Da Costa**
*Departamento de Física, Universidad Simón Bolivar*
*A.P. 89000, Caracas 1080-A, Venezuela*
e-mail: german.dacosta@gmail.com

## ABSTRACT

The free surface of hydrodynamic waves behaves as a time-varying planar curvilinear mirror, whose focal properties determine the light intensity distribution in a reflected light beam. Variational criteria for determination of foci of planar curvilinear mirrors illuminated by a coplanar light source are studied in the realm of Geometric Optics. Intrinsic functions of the optical setup (called focal potentials in the text) which are stationary at mirror points corresponding to cusp points of the caustic of reflected light rays are analyzed. The eccentricity of the osculating conic defined at each mirror point is shown to be a dimensionless, coordinate independent focal potential. An application to numerical analysis of light focusing by laser-illuminated hydrodynamic solitons is presented.

## I.INTRODUCTION

Smooth, light reflecting, cylindric surfaces are frequently met when hydrodynamic surface waves are studied by means of optical methods. An example is shown in fig.(1). It corresponds to an optical setup used for experimental analysis of hydrodynamic solitons [1]. Surface waves propagating in 1D along a horizontal water channel are illuminated by an inclined, collimated laser beam normal to the axis of the cylindric water surface. In any cross section normal to the cylinder axis the water surface profile behaves as a time-varying, mirror-like planar curve. Light intensity maxima appearing in the observation screen are due to concentration of light rays reflected from concave regions of the water surface. Their analysis furnishes valuable information on the time evolution of the surface profile. In other experiment (figs.2,3) the paraboloidal surface of a rotating heavy oil sample is further deformed by laser heating [2] thus allowing correction of axisymmetric surface aberrations. Any meridian section of the liquid surface is a planar curvilinear mirror of the kind cited above.

The aim of the present paper is to review variational principles allowing determination of planar mirror points such that the reflected light rays pass through light intensity maxima (light foci) in the reflected light beam. Generic planar optical systems are reviewed in Section (II). Variational methods are studied in Sections (III-V). Applications to optical analysis of hydrodynamic solitons are discussed in Section (VI). Conclusions are presented in Section (VII).

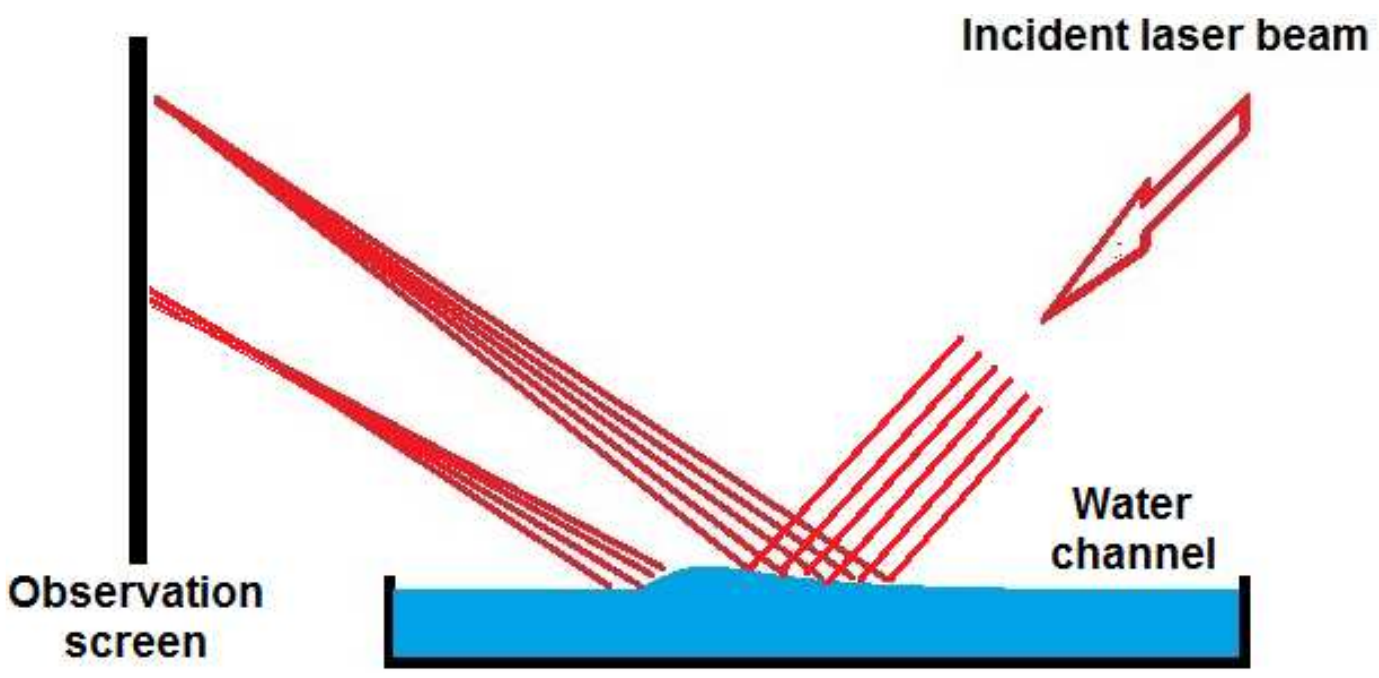

**Fig.(1):** *Laser light focusing by concave regions of a solitary water surface wave.*

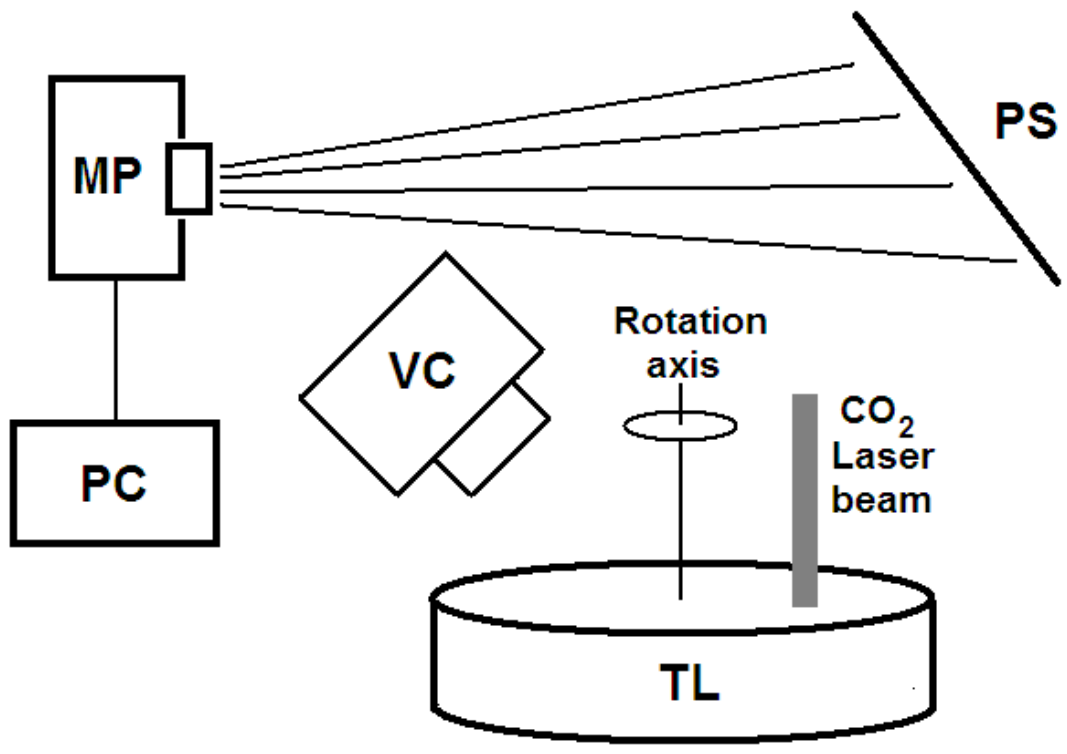

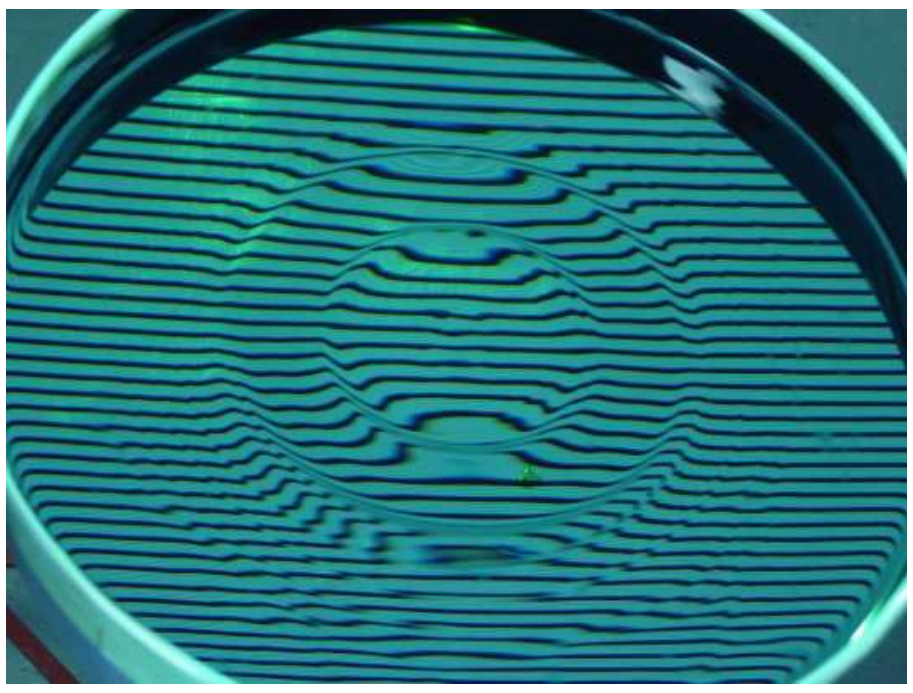

**Fig.(2)(left):** *Experimental setup used to record the axisymmetric 3D surface shape of a heavy oil sample deformed by rotation around a fixed vertical axis and (simultaneously) by heating with an off-axis laser beam. PC is a computer, MP a multimedia projector, PS a projection screen, TL the rotating liquid sample and VC a video camera.*
**Fig.(3)(right):** *Photographic record corresponding to fig.(2). A parallel fringe system is observed by reflection upon the free liquid surface, thus allowing optical coding and later numerical determination of the surface shape. All of the identical meridian sections of the liquid surface are planar curvilinear mirrors of the kind considered in the text.*

## II.GEOMETRIC PROPERTIES OF LIGHT RAYS REFLECTED BY CURVILINEAR PLANAR MIRRORS

Basic geometric properties of light rays reflected by curvilinear planar mirrors are reviewed in the present Section. The analysis does not take into account interference or diffraction effects. The branch of Mathematics underlying these optical properties is Differential Geometry in 2D Euclidean spaces [3-7]. Terms designing the same physical objects in Geometric Optics and Differential Geometry are compared and emphasized in bold letters in what follows. Fig.(4) represents a generic planar curvilinear mirror illuminated by light rays coming from a coplanar point light source (S). Point (P) is a current mirror point, (t, n) are straight lines respectively tangent and normal to the mirror at point (P) and (δ) is the incidence- reflection angle. Point (C) is **the contact point** of the light ray reflected at (P) with the **envelope** of the family of reflected light rays. This **envelope** is known in Optics as the **caustic** curve. The spatial density of reflected light rays is a maximum in its neighborhood. Point (V) is a **cusp** point of the **caustic**, currently called a **light focus** in Optics. Point ($P_V$) is the mirror point corresponding to (V). Analysis of general variational criteria allowing determination of points ($P_V$) is the main subject of the present paper. Point (W) is symmetric of point (S) with respect to tangent line (t). The loci of point (W) when point (P) travels along the mirror is known in Optics as the **reflected beam wavefront**. All of the reflected light rays are normal to the wavefront. In Differential Geometry texts the **wavefront** is called the **orthotomic** of the mirror with respect to point (S) [ref.4, page 92]. The **caustic** curve (**envelope** of reflected light rays) is also the **loci of curvature centers** (C) of the wavefront at the corresponding points (W). Following this property, the **caustic** is called the **wavefront evolute** in the language of Differential Geometry (ref.[4], page 33). Note that the **caustic** can also be considered as the **envelope** of lines normal to the **wavefront**. An intrinsic (coordinate-independent) analysis of the optical system is allowed by utilization of **pedal coordinates** (ref.[4], page 147) defined in eqs.(1,2).

$$r = |P - S| \qquad (1)$$

$$p = |T - S| \qquad (2)$$

Distances **(r, p)** as well as **κ, the mirror curvature,** are intrinsic properties of the optical system defined at each mirror point. The relation existing at each mirror point (P) between the **wavefront curvature ($\kappa_O$)** and **(r, p, κ)** is (ref.[4], page 147) is reminded in eq.(3).

$$\kappa_O = \frac{2r^2\kappa - p}{2r^3\kappa} \tag{3}$$

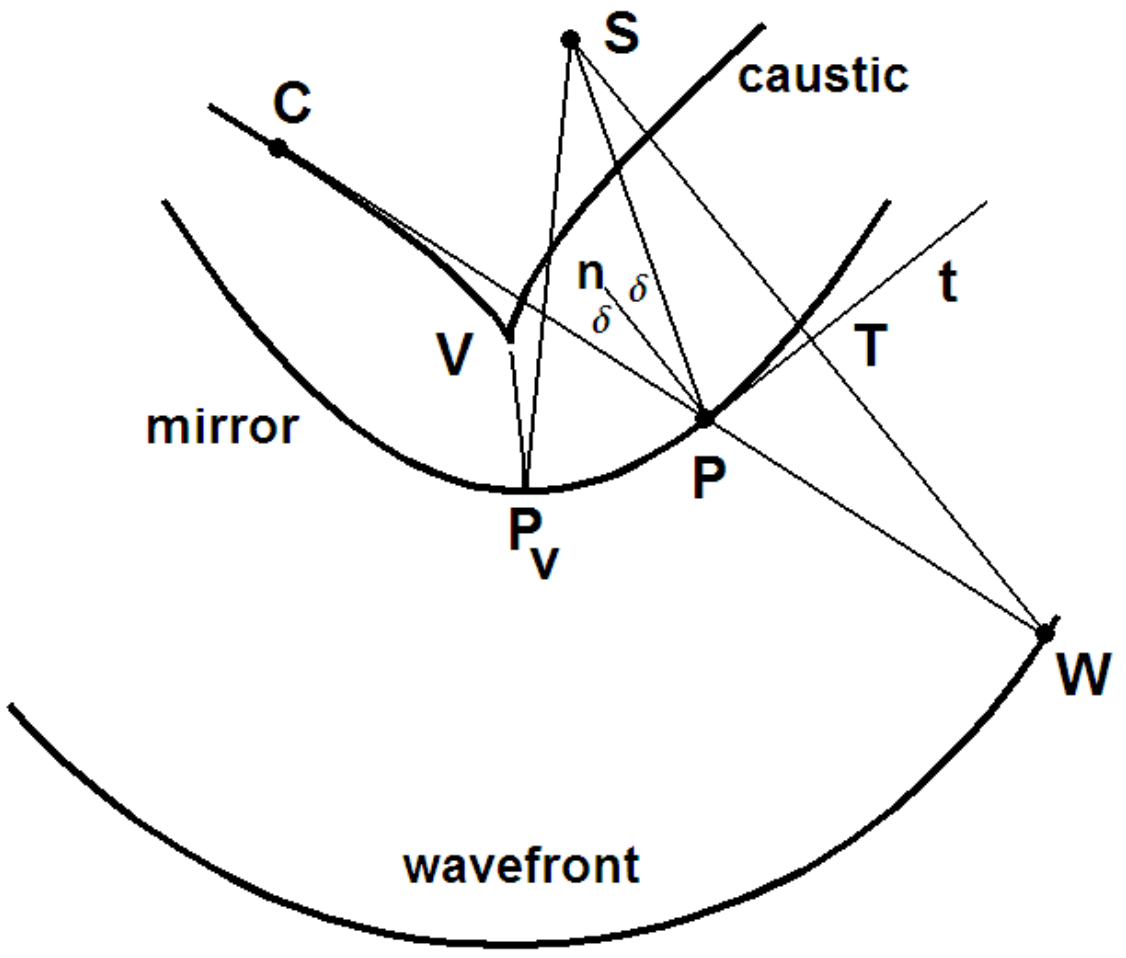

**Fig.(4).** *Mirror (M) is illuminated by light rays (SP) issued from the point light source (S). Point (W) is symmetric of (S) with respect to the tangent line (t). The light ray reflected at (P) is the prolongation of segment (WP). The wavefront of reflected light rays is the locus of (W) when (P) travels along the mirror. Straight line (WP) is normal to the wavefront. The caustic is the envelope of reflected light rays. Point (C) is the contact point of the light ray reflected at (P) with the caustic. Point (V) is a caustic cusp point, usually called a light focus in Optics. Point ($P_V$) is the mirror point corresponding to caustic point (V).*

It is assumed that in the considered region the mirror has no inflection points and consequently $(\kappa \neq 0)$. Taking into account that:

$$\cos(\delta) = p / r \tag{4}$$

eq.(3) also reads:

$$\kappa_O = \frac{2r\kappa - \cos(\delta)}{2r^2\kappa} \tag{5}$$

In eq.(5), **($\kappa_O$)** is written in terms of the local, intrinsic quantities **[r, κ, cos(δ)]**. At mirror points where [cos(δ) = 2rκ] the wavefront curvature ($\kappa_O$) is null. In that case point (W) (fig.4) is an inflection point of the wavefront.

## III.VARIATIONAL CRITERIA FOR DETERMINATION OF CAUSTIC CUSP POINTS

The mirror is usually described by parametric equations [3]:

$$x_P = x_P(s) \tag{6a}$$

$$y_P = y_P(s) \tag{6b}$$

where (s) is the arc length measured along the mirror and ($x_P$, $y_P$) are Cartesian coordinates of the generic mirror point (P). The corresponding caustic point (C) has equations [$x_C(s)$, $y_C(s)$]. At the particular caustic cusp point (V) both coordinates [$x_C(s)$, $y_C(s)$] have a relative extremum, for a caustic point (C) traveling along the caustic reverses its traveling sense when passing through (V) (eqs.7a,b).

$$dx_C(s)/ds = 0 \quad (7a)$$

$$dy_C(s)/ds = 0 \quad (7b)$$

The tangent vector to the caustic is thus null at (V). In above equations the s-derivatives are calculated at the value of arc length (s) corresponding to mirror point $P_V$ . Stationarity of $[x_C(s), y_C(s)]$ is therefore a variational criterion for determination of points ($P_V$). However, this criterion has the inconvenient of being coordinate-dependent. It is worthwhile to look for a criterion depending only on intrinsic properties of the optical system. It is shown (ref.[4], pages 18, 122, 128, 144) that the evolute of a curve has a cusp when the curve has a vertex, the latter cited being defined as a curve point where the first derivative of curvature is zero. In our actual case, and in the language of Geometric Optics, this means that the wavefront curvature ($\kappa_O$) is stationary at wavefront points (W) corresponding to caustic cusp points (V), thus at mirror points ($P_V$). Therefore **an intrinsic criterion for determination of caustic cusp points** is:

$$d\kappa_O(s)/ds = 0 \quad (8)$$

Solutions of eq.(8) yield the arc length values corresponding to points ($P_V$). It is possible to use distance (r) instead of arc length (s) as the parameter varying along the mirror whenever function r(s) is biunivoque, which requires $(dr/ds \neq 0)$ in the considered interval of values of (s). In what follows derivatives with respect to variable (r) will be named by primes. The condition (eq.8) for existence of a cusp point is then written in terms of parameter (r):

$$\frac{d\kappa_O(s)}{ds} = \frac{d\kappa_O(r)}{dr}\frac{dr}{ds} = \kappa_O'(r)\frac{dr}{ds} = 0 \quad (9)$$

We conclude that any mirror point such that (dr/ds = 0) or $[\kappa_O'(r) = 0]$ is a ($P_V$) point. Note that the same conclusion is valid for any function $\alpha(r)$ such that $\alpha'(r) = \beta(r)\kappa'_O(r)$, where $\beta(r)$ is any arbitrary function of (r) which has no roots in the considered interval of values of (r). Integration by parts yields:

$$\alpha = \beta\kappa_O - \int \beta'\kappa_O dr + Cte \quad (10)$$

where Cte is an arbitrary constant. As $\beta(r)$ can be arbitrarily chosen eq.(10) represents an infinite family of functions which are stationary at ($P_V$) mirror points. Whenever the indefinite integral in eq. (10) can be formally integrated we get a finite expression for the new function $\alpha(r)$. In this search for new members of the family it is worthwhile to calculate explicitly the derivative $\kappa_O'(r)$ , which requires using the property (ref.[4], page 147) reminded in eq.(11).

$$p' = r\kappa \quad (11)$$

From eqs.(3, 11) we thus get:

$$\kappa_O' = -\frac{1}{2\kappa^2 r^4}\left(3\kappa^2 r^2 - pr\kappa' - 3p\kappa\right) \quad (12)$$

The condition $[\kappa_O'(r) = 0]$ is therefore equivalent to:

$$3\kappa^2 r^2 - pr\kappa' - 3p\kappa = 0 \quad (13)$$

Note that eq.(13) contains only quantities (r, p, κ) pertaining directly to the mirror, not to the wavefront. Now we want to formulate this condition in terms involving only distances (r, p). The derivative **(κ')** is calculated from eq.(11):

$$\kappa' = -\frac{1}{r^2} p' + \frac{1}{r} p'' \tag{14}$$

which replaced in eq.(13) leads to:

$$\frac{p''}{p'} - 3\frac{p'}{p} + \frac{2}{r} = 0 \tag{15}$$

By integration of the left side of eq.(15) we conclude that:

$$\frac{d}{dr}\left[ Log\left( \frac{p' r^2}{p^3} \right) \right] = 0 \tag{16}$$

or equivalently:

$$\left( \frac{p' r^2}{p^3} \right)' = 0 \tag{17}$$

Replacing eq.(11) into eq.(17) yields:

$$\left( \frac{\kappa r^3}{p^3} \right)' = 0 \tag{18}$$

which means that:

$$\frac{d}{dr}\left[ \frac{1}{\kappa} \cos^3(\delta) \right] = \frac{d}{dr}\left[ R\cos^3(\delta) \right] = 0 \tag{19}$$

where (R = 1/κ) is the mirror curvature radius. Eq.(19) implies that function:

$$\phi = R\cos^3(\delta) = \frac{1}{\kappa}\left( \frac{p}{r} \right)^3 \tag{20}$$

is (as well as $\kappa_O$) stationary at caustic cusp points. As reminded in Section (I), a similar result corresponding to the particular case when the point light source is placed in the far field of the mirror (collimated beam illumination) was presented in refs.[1,8]). By differentiation of eq. (20) we conclude that :

$$\phi' = \frac{p^2}{\kappa^2 r^4}\left( 3\kappa^2 r^2 - pr\kappa' - 3p\kappa \right) \tag{21}$$

and consequently (eqs.12, 21) that:

$$\phi' = -2p^2 \kappa_O' \tag{22}$$

With the choice ($\beta$ = - $2p^2$) eq.(22) shows that ($\phi$) belongs to the family of functions defined by eq.(10), which are called **focal potentials** in what follows. They all share the basic property of being stationary at mirror points ($P_V$) corresponding to caustic cusp points (V). It will be shown in Sections (III,IV) that($\phi$, $\kappa_O$) are privileged members of the family of focal potentials, for they admit a sound geometric interpretation in terms of the lengths of the axes of elliptical mirrors closely ressembling the given mirror at each point (P). Even in the present frame, algebraic combination of ($\phi$, $\kappa_O$) yields new focal potentials with interesting physical properties. Note that the r-derivative of any expression of the kind F($\phi$,$\kappa_O$) (where F designs an arbitrary function of the only variables $\phi$,$\kappa_O$) is itself null whenever ($\phi'$ = $\kappa_O'$ = 0), for:

$$\frac{dF(\phi,\kappa_O)}{dr} = \frac{\partial F(\phi,\kappa_O)}{\partial \phi}\phi'(r) + \frac{\partial F(\phi,\kappa_O)}{\partial \kappa_O}\kappa_O'(r) \tag{23}$$

In particular the function ( F($\phi$,$\kappa_O$) = $\phi\kappa_O$ ) has the important property of being dimensionless. With the definition:

$$\lambda = \frac{\cos(\delta)}{r\kappa} = \frac{R\cos(\delta)}{r} \tag{24a}$$

we get from eqs.(5,20):

$$\gamma = 2\phi\kappa_O = \lambda(2-\lambda)\cos^2(\delta) \tag{24b}$$

which represents the dimensionless focal potential $(\gamma)$ as a function of two independent dimensionless parameters $[\lambda, \cos(\delta)]$. Note the geometric meaning of these parameters: **a)** $\cos(\delta)$ is the cosine of the incidence-reflection angle of light rays; **b)** $(\lambda)$ is the ratio between the projection [Rcos($\delta$)] of the mirror curvature radius (R) upon the incident light ray, and the distance (r) from the mirror point to the light source. The case ($\lambda$ = 2) is seen from eq.(5) to correspond to an inflection point of the wavefront, which gives rise to a caustic asymptote $(\kappa_O \to 0, R_O \to \infty)$. In Section (V) it is shown that focal potential $(\gamma)$ depends only on the osculating ellipse eccentricity, which is thus itself a dimensionless focal potential.

As a final remark to the present Section let us note that while stationarity of the wavefront curvature ($\kappa_O$ ) has a clear physical interpretation as being responsible for the maximum light concentration existent at caustics cusp points, a physical interpretation for ($\phi$) and other focal potentials is still lacking. The geometric viewpoint (inspired essentially in ref[4] ) adopted in forecoming Sections (IV - V) gives rise to an interesting physical interpretation for all of the focal potentials defined above.

## IV. PARTICULAR CASE : ELLIPTIC MIRRORS

A particular case of above presented general properties is shown in fig.(5), where the given mirror (E) is elliptical and the light source is placed at the ellipse focus ($F_1$). A well known property of elliptic mirrors states that any light ray issued from ($F_1$) and later reflected at the mirror surface passes through the other focus ($F_2$). The caustic by reflection then reduces to one single point ($F_2$) and all of the mirror points are of the kind named ($P_V$) in fig.(4). The wavefront curvature radius ($R_{OE}$ = $1/\kappa_{OE}$ , where subindex "E" emphasizes the fact that we are dealing with an elliptic mirror) must therefore be stationary at any mirror point. This implies that ($R_{OE}$) must be a constant quantity, which is calculated in what follows. Let us call (2a, 2b) the major and minor ellipse axes respectively. Following the geometric construction depicted in fig.(4) we note (fig.5) that point (W) (symmetric of point $F_1$ with respect to the tangent line at P) pertains to the wavefront of reflected light rays, and that the curvature center of the wavefront at (W) is the contact point of reflected light ray (WP) with the caustic. Consequently, the length of segment ($WF_2$) equals the curvature radius of the ellipse wavefront at (W) and we have:

$$WF_2 = WP + PF_2 = F_1P + PF_2 = 2a \tag{25}$$

The wavefront is thus a circle with radius ($R_{OE}$ = 2a) centered at focus ($F_2$).

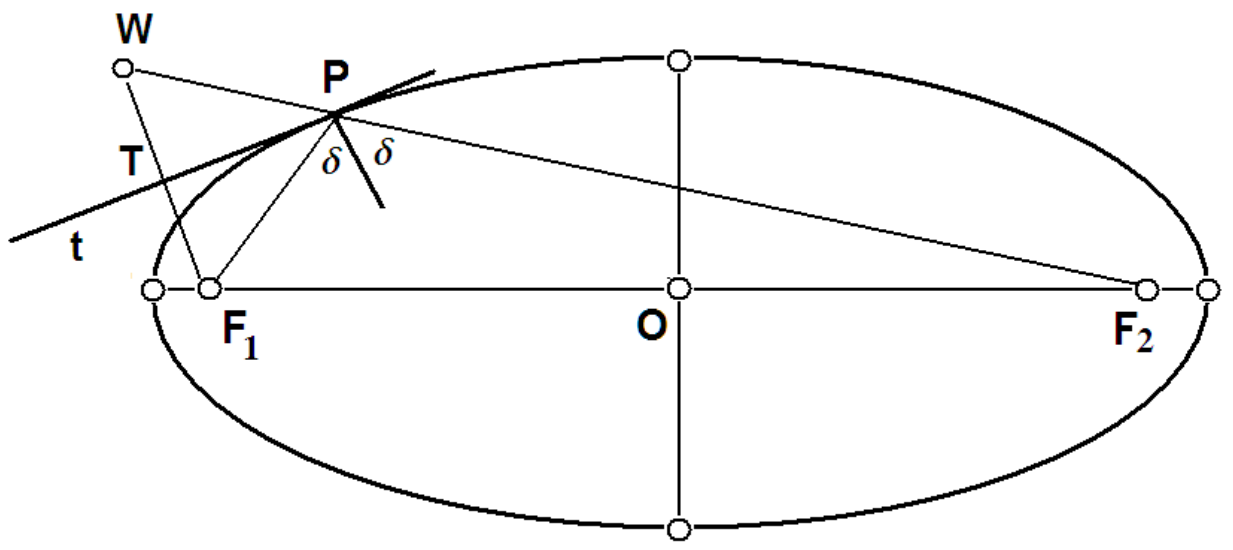

**Fig.(5).** *Above picture is a particular case of fig.(4) when the mirror is elliptic and the point light source (S) coincides with one of the ellipse foci ($F_1$). Line (t) is tangent to the ellipse at point (P) and point (W) is symmetric of point (P) with respect to (t). Consequently (W) pertains to the wavefront of reflected light rays and the length of segment $WF_2$ equals the wavefront curvature radius. All of the reflected light rays pass through the other focus ($F_2$ ).*

It is also interesting to calculate the constant value of focal potential **(**$\phi_E$**)**. The position vector $\overline{OP}(\tau)$ of a current mirror point (P) in the rectangular coordinate system (Oxy) (fig.6) has coordinates:

$$\overline{OP}(\tau) = [x(\tau), y(\tau)] \tag{26}$$

with:

$$x(\tau) = a\cos(\tau) \tag{27a}$$

$$y(\tau) = b\sin(\tau) \tag{27b}$$

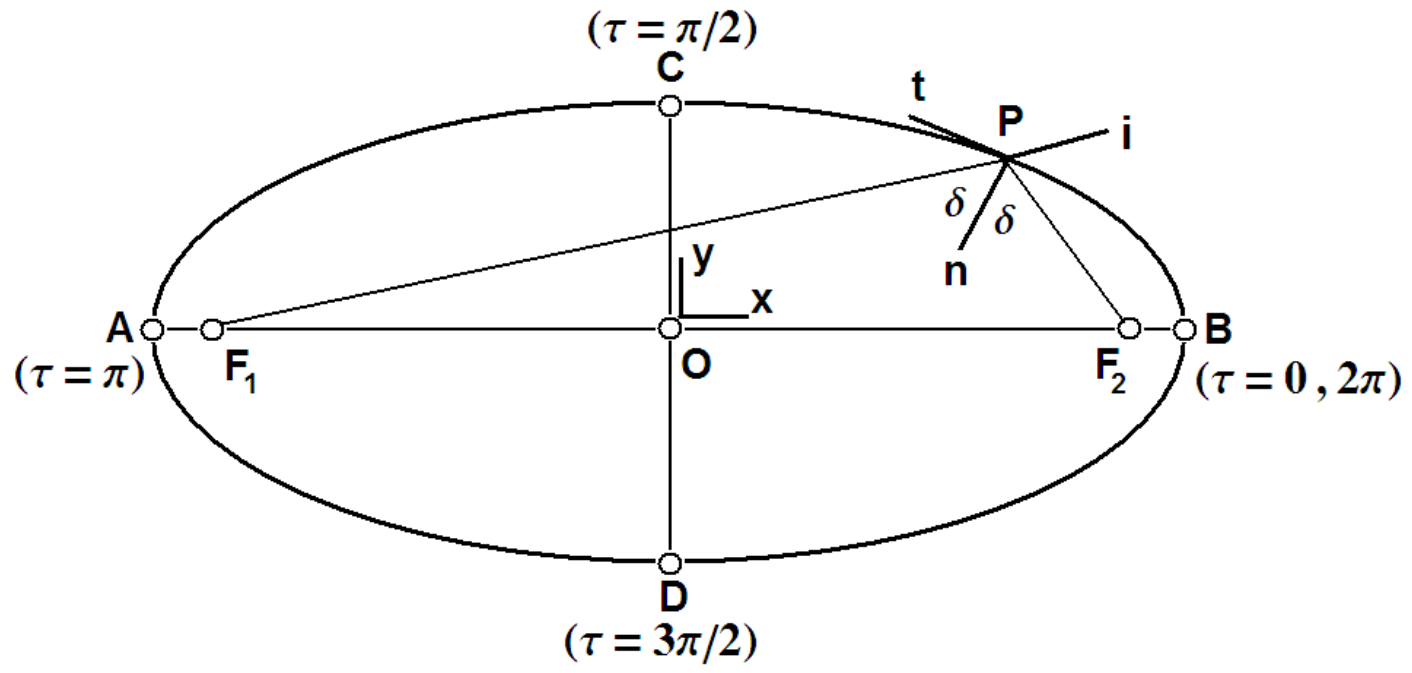

**Fig.(6).** *Elliptic mirror (foci $F_1$,$F_2$) illuminated by light rays issued from a point light source placed at ($F_1$). In above picture (t,n) are the unit vectors respectively tangent and normal to the mirror at a current point (P), (i) is the unit vector of the light ray incident at (P), (Oxy) is a rectangular coordinate axis, (AB, CD) are the ellipse axes and ($\tau$) is a parameter varying along the mirror.*

In above equations $(\tau)$ is a real-valued parameter ($0 \le \tau \le 2\pi$). The parameter value $(\tau = 0)$ corresponds to vertex (B) of the major ellipse axis. The parameter value increases when point (P) travels along the ellipse in counter-clockwise sense. Note that τ(s) is a monotonically increasing (thus biunivoque) function of the ellipse arclength (s). By differentiation of eqs.(27a,b) we get the unit vector tangent to the mirror at point (P):

$$\bar{t}(\tau) = \frac{[\dot{x}(\tau), \dot{y}(\tau)]}{\sqrt{\dot{x}^2(\tau) + \dot{y}^2(\tau)}} = \frac{[-a\sin(\tau), b\cos(\tau)]}{\sqrt{a^2\sin^2(t) + b^2\cos^2(t)}} \qquad (28)$$

where points upon functions stand for derivatives with respect to **(τ)**. The unit normal vector is [3]:

$$\bar{n}(\tau) = J\,\bar{t}(\tau) \qquad (29)$$

where **(J)** is the rotation operator turning the adjacent vector (π/2) radians in counter-clockwise sense. Therefore:

$$\bar{n}(\tau) = \frac{[-\dot{y}(\tau), \dot{x}(\tau)]}{\sqrt{\dot{x}^2(\tau) + \dot{y}^2(\tau)}} = \frac{[-b\cos(\tau), -a\sin(\tau)]}{\sqrt{a^2\sin^2(t) + b^2\cos^2(t)}} \qquad (30)$$

The coordinates of the position vector of focus ($F_1$) are:

$$\overline{OF_1} = [-\sqrt{a^2 - b^2}, 0] \qquad (31)$$

The unit vector of  the incident light beam is:

$$\bar{i}(\tau) = \frac{\overline{OP} - \overline{OF_1}}{\sqrt{(\overline{OP} - \overline{OF_1}).(\overline{OP} - \overline{OF_1})}} \qquad (32)$$

A straightforward calculation yields:

$$\overline{i(\tau)}.\overline{n(\tau)} = \frac{-b}{\sqrt{a^2\sin^2(\tau) + b^2\cos^2(\tau)}} \qquad (33)$$

where the dot means scalar product of adjacent vectors. On the other hand, the mirror curvature radius is given by [3]:

$$R(\tau) = \frac{[\dot{x}^2(\tau) + \dot{y}^2(\tau)]^{3/2}}{\ddot{y}(\tau)\dot{x}(\tau) - \dot{y}(\tau)\ddot{x}(\tau)} \qquad (34)$$

Consequently:

$$R(\tau) = \frac{[a^2\sin^2(\tau) + b^2\cos^2(\tau)]^{3/2}}{ab} \qquad (35)$$

By comparison of eqs.(33,35) we conclude that the scalar quantity:

$$\alpha = R(\tau)[\bar{i}(\tau).\bar{n}(\tau)]^3 = -b^2 / a \qquad (36)$$

remains constant when point P(τ) travels along the elliptic mirror. Note that the angle between unit vectors $(\bar{i},\bar{n})$ is (π - δ) radians. So we have:

$$\bar{i}(\tau).\bar{n}(\tau) = -\cos[\delta(\tau)] \qquad (37)$$

We conclude that the quantity:

$$\phi_E = R(\tau)\cos^3[\delta(\tau)] = b^2 / a = R(0) = R(\pi) \qquad (38)$$

remains constant when the considered point travels along the mirror. Both quantities ($\phi_E$ = $b^2$/a, $R_O$ =2a) are thus independent of the mirror parameter (τ). Note that ($\phi_E$) is just the generic focal potential (Φ) defined in Section (II) calculated in the particular optical system treated in the present Section. In the latter cited case (Φ) is equal to the maximum value of the ellipse curvature radius, which is attained at vertex (A,B) (fig.6). A generalization of this property for generic optical systems of the kind depicted in fig.(4) is studied in Section (V).

## V.GEOMETRIC INTERPRETATION OF FOCAL POTENTIALS

The geometric viewpoint adopted in what follows, assigning an osculating conic to each point (P) of mirror (M), is thoroughly emphasized in refs.[4-7]. The conic is an ellipse if point (S) lies inside the circle of curvature of the wavefront at (W) (that is, the circle centered at the wavefront curvature center C and tangent to the wavefront at point W) or a hyperbola if it lies outside the circle (ref.[4], page 142). Forecoming figures of the present text are drawn assuming that the osculating conic is an ellipse. However, conclusions of the corresponding analysis are also valid in the hyperbolic case, for they are based on universally valid formulae introduced in preceding Sections (II, III). Conclusions of Section (IV) suggest that it would be worthwhile to define at each mirror point $(P \in M)$ an ellipse (E) (fig.7) with one focus (S) coincident with the light source and locally ressembling (M) as closely as possible. It seems reasonable to expect that focal potentials of (M) should be related to the axes of the associated ellipse (E), same as ($R_{OE}$, $\phi_E$) are related to the elliptic mirror axes lengths in Section (IV). In this aim, let us draw (fig.7) through each mirror point (P) the ellipse (E) (called the osculating ellipse of mirror M at point P) fulfilling the following conditions [ref.4, pages 141-142]: **a)** one of its foci coincides with the light source (S), **b)** it is tangent to the mirror at point (P), **c)** it has at point (P) the same curvature radius as the mirror (M). Note that there is one and only one solution to this problem. For in general an ellipse is determined by six parameters, say the Cartesian coordinates of both foci and of an arbitrary ellipse point [3]. In the present case the six given parameters are the coordinates of points (P,S) and the common slope and curvature of (M,E) at point (P).

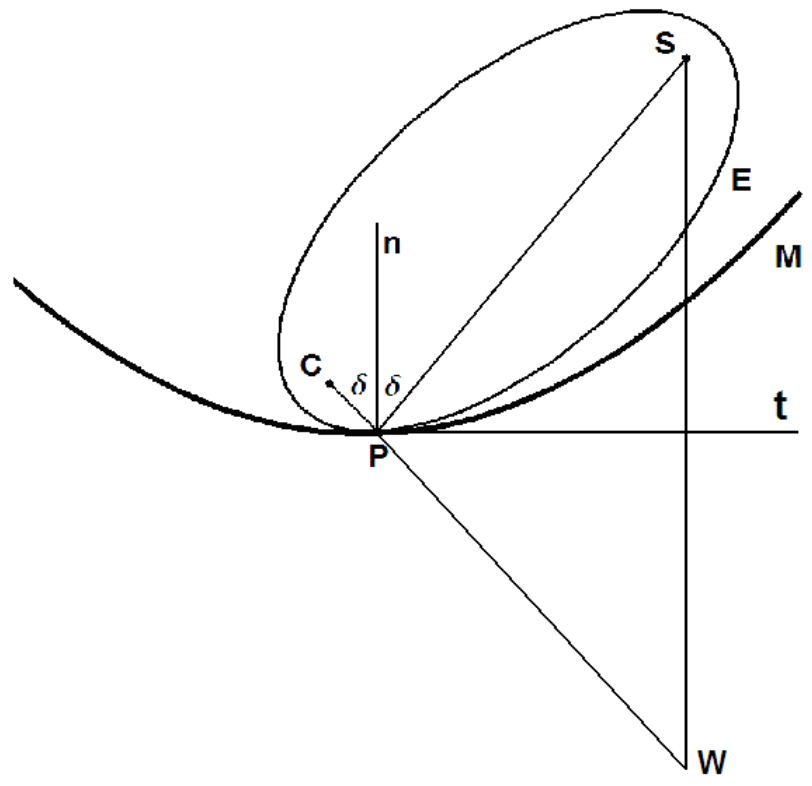

**Fig.(7):** *Osculating ellipse (E) of mirror (M) at point (P).*

Focus (C) coincides with the contact point of the light ray reflected at (P) with the envelope of light rays reflected at (M). The following properties are worth noting:

**a)**All of the osculating ellipse geometric properties (such as the location of foci and the axes lengths) are univocally determined once the position of point (P) is given.

**b)**The focus of (E) other than (S) coincides with the contact point (C) of the light ray reflected at (P) with the caustic of (M). For the geometric construction (sketched at fig.4) of contact point (C) is the same for both (M,E) (that is: with origin at W take along the reflected ray a segment with length $R_O = R_{OE}$ ) and leads therefore in both cases to the same point (C).

**c)**If we consider (E) as a mirror with a focus coincident with the light source (S), we remind (fig.6) that ($R_{OE}$ = 2a), where now (2a) is the osculating ellipse major axis. On the other hand, eq.(5) shows that for any arbitrary mirror the wavefront curvature depends only on the local mirror curvature (κ), on the light ray incidence angle (δ) and on the distance (r) to the light source. As all three quantities are shared by (M) and (E) at any point (P), we conclude that ($R_O = R_{OE}$) at any mirror point (P). Consequently we have ($R_O = R_{OE}$ = 2a) at any mirror point (P), in particular at points ($P_V$). And as ($R_O$) is stationary at mirror points ($P_V$), the same is true for (2a). In addition, the distance $|C-S|$ between the ellipse foci is:

$$\sqrt{x_C{}^2 + y_C{}^2} = 2\sqrt{a^2 - b^2} \tag{39}$$

and consequently:

$$b^2 = a^2 - (x_C{}^2 + y_C{}^2)/4\,. \tag{40}$$

At points ($P_V$) all three quantities (a, $x_C$ , $y_C$) are stationary (eqs.7a,b) and we conclude (eq.40) that also the ellipse minor axis length (2b) is stationary at points ($P_V$). **We thus arrive at the important conclusion that the semiaxes (a,b) are focal potentials of the optical system. The whole geometric structure of the osculating ellipse (location of foci, axes lengths, eccentricity) is therefore stationary at mirror points ($P_V$).**

**d)**Focal potentials (a,b) are calculated as functions of (r, p, κ) as follows. Eq.(3) yields:

$$a = \frac{R_O}{2} = \frac{r^3\kappa}{2r^2\kappa - p} \tag{41}$$

In triangle CSP we have (fig.7):

$$|C-S|^2 = |C-P|^2 + |P-S|^2 - 2|C-P||P-S|\cos(2\delta) \tag{42}$$

which using (eqs.1,4,39) is term by term re-written as:

$$4(a^2 - b^2) = (R_O - r)^2 + r^2 - 2(R_O - r)r(2(p/r)^2 - 1) \tag{43}$$

Replacing ($R_O$, a) given by eqs.(4,41) respectively into eq.(43) we find:

$$b^2 = \frac{p^3}{2r^2\kappa - p} \tag{44}$$

From eqs.(41, 44) we also get:

$$\frac{b^2}{a} = \frac{1}{\kappa}\left(\frac{p}{r}\right)^3 = \phi \tag{45}$$

Eq.(45) reinforces the geometric interpretation of focal potential (ϕ) already pointed in preceding Sections:

**a)In an elliptic mirror (E)** with a point light source placed at one of its foci ($F_1$) all of the reflected light rays pass through the other focus ($F_2$), which is the (degenerate) caustic cusp point (V). The quantity $\phi = (1/\kappa)(p/r)^3$ has the constant value $(b^2/a)$, thus being stationary all along (E).

**b)In an arbitrarily shaped mirror (M)** the same quantity is stationary only at mirror points ($P_V$) corresponding to the caustic cusp points (V). In the first case the quantity $(\phi = b^2/a)$ is related to the elliptic mirror axes, while in the second case the same relation stands relative to the osculating ellipse axes.

Eqs.(41,44, 24a,b) also yield:

$$\left(\frac{b}{a}\right)^2 = \phi\kappa_O = \gamma/2 \tag{46}$$

This gives a geometrical interpretation for the dimensionless focal potential (γ) as being directly related to the eccentricity $(\varepsilon)$ of the osculating ellipse:

$$\gamma = 2\left(\frac{b}{a}\right)^2 = 2(1-\varepsilon^2) \tag{47}$$

Also inverting the relations :

$$\kappa_O = \frac{1}{2a} \tag{48a}$$

$$\phi = \frac{b^2}{a} \tag{48b}$$

we get:

$$a = \frac{1}{2\kappa_O} \tag{49a}$$

$$b = \sqrt{\frac{\phi}{2\kappa_O}} \tag{49b}$$

Differentiation of eqs.(48a,b) yields:

$$\kappa_O' = -\frac{a'}{2a^2} \tag{50a}$$

$$\phi' = \frac{2bb'a - a'b^2}{a^2} \tag{50b}$$

which combined with eq.(22) yields:

$$b' = \frac{b^2 + p^2}{2ab} a' \tag{51}$$

We thus conclude that the role played by focal potentials ($\phi$, $\kappa_O$ ) on one hand and axes lengths (a, b) on the other hand is totally symmetric. As well as ($\phi' = 0$) if and only if ($\kappa_O' = 0$) the necessary and sufficient condition to have ($a' = 0$) is ($b' = 0$). Any other focal potential is derived from the axes lengths (a,b) by means of relations similar to eq.(10) or by arbitrary functional combinations of the kind F(a,b).

## VI.NUMERICAL APPLICATION TO HYDRODYNAMIC SOLITON ANALYSIS

As an example, preceding theory is used for rapid numerical determination of the points in a laser illuminated hydrodynamic soliton giving rise to focal points in the reflected light beam. The equation of a hydrodynamic soliton propagating along a rectilinear water channel is [1]:

$$f[x,t] = A.Sech^2[\frac{1}{2h}\sqrt{\frac{3A}{h}}\left(x - \sqrt{gh}\left(1 + \frac{A}{2h}\right)t\right)] \tag{52}$$

In eq.(52) , f[x,t] is the soliton profile above the water free surface at rest¸ (A) is the maximum soliton height, (h) is the water depth at rest , (g) is the acceleration of gravity, (x) is the abscissa along the water channel and (t) is propagation time. The soliton profile is represented in fig.(8) for the particular numerical values: A = 1cm , h = 10cm , g = 981 cm/s$^2$ , t = 0 . The focal potential ($\phi$) studied in preceding Sections, now calculated in Cartesian coordinates, is given in eq.(53). In eq.(53) (m) is the slope of the collimated laser beam incident upon the soliton profile from the right side of the picture. Fig.(9) represents the numerator of the first derivative of $\phi(x)$ for the same numerical values and (m = 0.2).

$$\phi(x) = (m - f'(x))^3 / f''(x) \tag{53}$$

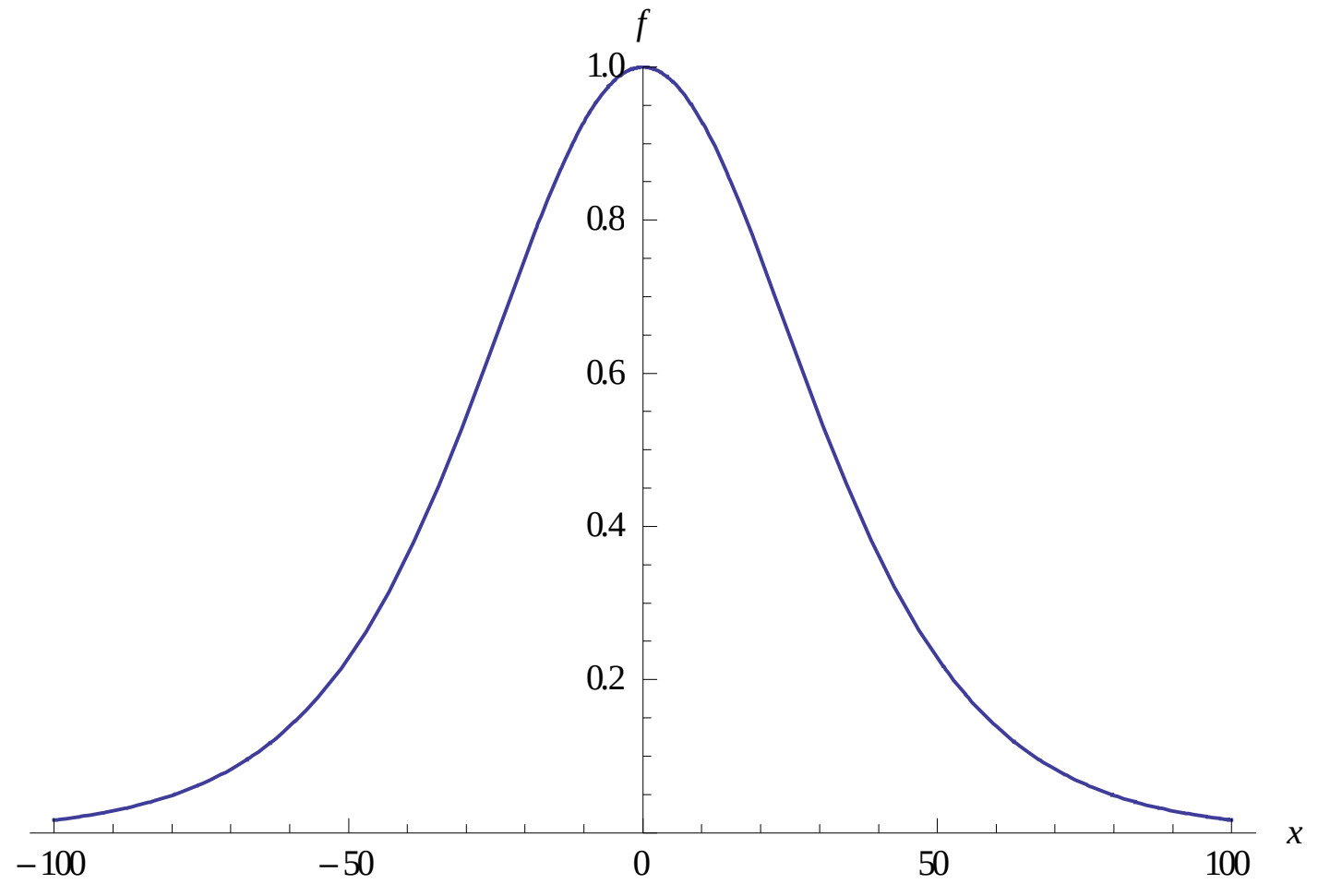

*Fig.8: Profile of a hydrodynamic soliton propagating in a water channel*

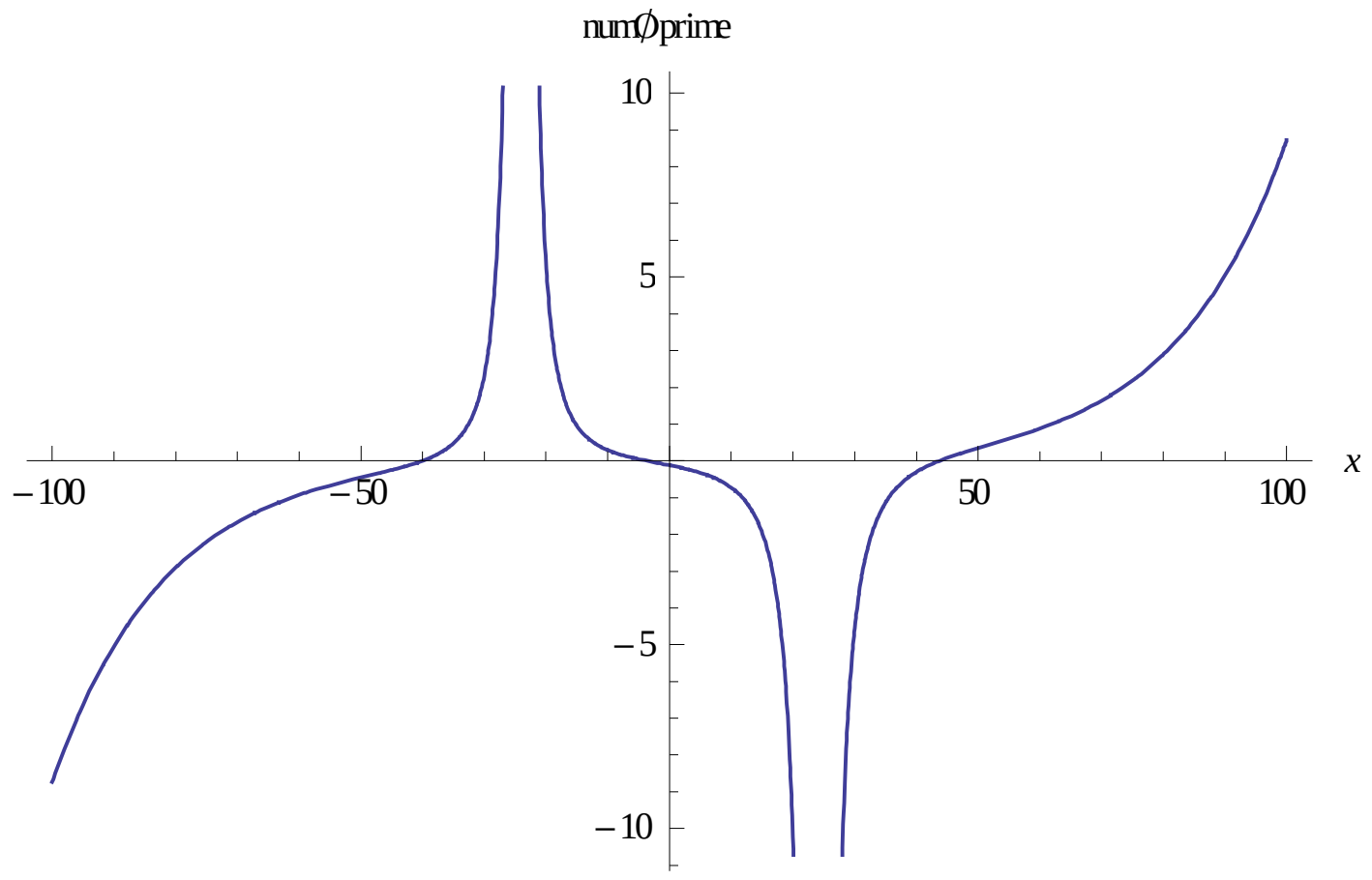

*Fig.9: Vertical axis represents the numerator of* $\phi'(x)$ *, where* $\phi(x)$ *is the focal potential of the solitonic profile in fig. (8).*

Direct visual inspection of fig.(9) shows that the roots of $\phi'(x)=0$ have approximate values : $(x_1 \approx -40cm, x_2 \approx -5cm, x_3 \approx 45cm)$ which are used as seeds for its more precise numerical calculation, thus obtaining: $(x_1 \approx -40{,}0193cm, x_2 \approx -3.6578cm, x_3 \approx 43.7616cm)$. The whole procedure is easily performed using *Mathematica* software. It involves first the symbolic calculation of $\phi'(x)$ and then the numerical solution of equation $\phi'(x)=0$.

## VII.CONCLUSIONS

**1)** Previous experiments with hydrodynamic solitons illuminated by a collimated laser beam [1, 8] lead to the conclusion that the water surface profile behaves as a 2D planar curvilinear mirror. The quantity $[\phi = \cos^3(\delta)/\kappa]$ is shown to be stationary at mirror points (called $P_V$ in the text) giving rise to light foci (called V) in the reflected light beam. In the particular case of collimated illumination (arbitrary slope *m*) the expression for (ϕ) in Cartesian coordinates reduces to eq.(53). The derivative $\phi'(x)$ is null at mirror points ($P_V$). In the present text the stationarity property is shown to be valid not only for quantity $(\phi)$ under collimated illumination, but also for any arbitrary position of a point light source (S), and for a whole class of functions called **focal potentials** of the optical setup. Quantity $(\phi)$ is just one of them. The procedure is as follows.

**2)** Basic properties of planar optical systems formed by a generic curvilinear mirror and a point light source are reminded in Section (II). Among them, stationarity of the wavefront curvature ($\kappa_O$) is a classic variational criterion for determination of mirror points ($P_V$) corresponding to caustic cusp points (V). So we have at least two basic focal potentials, namely $(\phi, \kappa_O)$.

**3)** An infinite family of focal potentials additional to $(\phi, \kappa_O)$ and sharing the same stationarity property is studied in Section (III). Focal potentials depend in general on three intrinsic properties of the optical system at each mirror point: **a)** the local mirror curvature (κ), **b)** the distance (r) from the considered mirror point to the light source, and **c)** the incidence-reflection angle (δ). However, focal potential named $[\phi = \cos^3(\delta)/\kappa]$ depends only on (κ, δ) and has a particularly simple algebraic structure. The product ($\kappa_O\phi$) is also a focal potential which depends on (κ, r, δ) but has the conceptual interest of being dimensionless.

**4)** A geometric interpretation of focal potentials in terms of the osculating ellipse associated to each mirror point is given in Sections (IV-V). It is shown that focal potentials $(\phi, \kappa_O)$ are algebraically related to the osculating ellipse axes lengths. Quantity $(\kappa_O\phi)$ is related to the osculating ellipse eccentricity, which is itself a dimensionless, coordinate independent focal potential.

**5)** An application to rapid numerical determination of points of a hydrodynamic soliton giving rise to focal points in the reflected light beam is presented in Section (VI). In general terms the same procedure allows theoretical modeling of any unknown water surface profile and later comparing its caustics (calculated by above methods) with experimental records of the actual reflected light field.

## ACKNOWLEDGMENT

The author acknowledges support of Physics Departments of Univ. Simón Bolivar (USB, Caracas, Venezuela) and Facultad de Ingeniería, Universidad de la República (UdelaR, Montevideo, Uruguay).

----------------------------------------------------
----------------------------------------------------